\documentclass[showpacs,prl,floatfix,twocolumn,amsmath,a4]{revtex4}

\usepackage{graphicx}

\begin{document} 

\title{KCrF$_3$: Electronic Structure, Magnetic and Orbital Ordering from First Principles}



\author{Gianluca Giovannetti$^{1,2}$, Serena Margadonna$^{3}$, Jeroen van den Brink$^{1,4}$}

\address{
$^1$Institute Lorentz for Theoretical Physics, Leiden University, 
          P.O. Box 9506, 2300 RA Leiden, The Netherlands\\ 
$^2$Faculty of Science and Technology and MESA+ Research Institute, University of Twente,
            P.O. Box 217, 7500 AE Enschede, The Netherlands\\ 
$^3$School of Chemistry, University of Edinburgh, 
West Mains Road, Edinburgh EH9 3JJ, UK\\ 
$^4$Institute for Molecules and Materials, Radboud Universiteit Nijmegen,
P.O. Box 9010, 6500 GL Nijmegen, The Netherlands.}

\begin{abstract} 

The electronic, magnetic and orbital structures of KCrF$_3$ are determined in all its recently identified crystallographic phases (cubic, tetragonal, 
and monoclinic) with a set of {\it ab initio} LSDA and LSDA+U calculations. The high-temperature undistorted cubic phase is metallic within the LSDA, 
but at the LSDA+U level it is a Mott insulator with a gap of 1.72 eV. The tetragonal and monoclinic phases of KCrF$_3$ exhibit cooperative Jahn-Teller 
distortions concomitant with staggered $3x^2-r^2$/$3y^2-r^2$ orbital order. We find that the energy gain due to the Jahn-Teller distortion 
is 82/104 meV per chromium ion in the tetragonal/monoclinic phase, respectively. These phases show A-type magnetic ordering and have a bandgap of 2.48 eV.  In this Mott insulating state KCrF$_3$ has a substantial conduction bandwidth of 2.1 eV, leading to the possibility for the kinetic energy of charge carriers in electron- or hole-doped derivatives of KCrF$_3$ to overcome the polaron localization at low temperatures, in analogy with the situation encountered in the colossal magnetoresistive manganites.

\end{abstract}

\date{\today} 


\maketitle

\section{Introduction}

About a decade ago, the discovery of the Colossal Magnetoresistance (CMR) effect in doped manganites caused a surge of interest in these perovskite 
oxides~\cite{Jin94,Imada98}. The particular physical properties of the CMR materials are related to the fact that their parent compound LaMnO$_3$ 
contains Mn$^{3+}$ ions with 4 electrons in its $d$-shell. On the one hand, the presence of these Jahn-Teller active ions leads to a strong coupling 
between the electrons and the lattice, giving rise to polaron formation which is widely perceived to be essential for the CMR 
effect~\cite{Millis95,Mannella05}. On the other hand, when doped, the $d^4$ high spin state leads via the double exchange mechanism to a 
ferromagnetic metallic state with a large magnetic moment making the system easily susceptible to externally applied magnetic fields~\cite{Zener51}. 
Moreover the presence of strong electron correlations and an orbital degree of freedom, to which the Jahn-Teller effect is directly related, leads 
to an extraordinary rich phase diagram at higher doping concentration, displaying a wealth of spin, charge, orbital and magnetically ordered 
phases~\cite{Brink04a,Brink99a}. It follows that the high spin $d^4$ state of Mn$^{3+}$ is intimately related to a plethora of physical phases, 
effects and properties. However, it is important to note that the high spin $d^4$ state is {\it not} exclusive to trivalent manganese.

Formally high spin Cr$^{2+}$ is electronically equivalent to Mn$^{3+}$. However, divalent chromium due to its low ionization potential 
is rarely found in solid state systems, making KCrF$_3$ an intriguing example. Recently, we characterized in detail 
the temperature-dependent crystallographic phase diagram of KCrF$_3$ revealing strong structural, electronic, and magnetic similarities 
with LaMnO$_3$~\cite{Margadonna,Margadonna2}, including the presence of Jahn-Teller distortions and 
orbital ordering, and orbital melting at high temperature. Here we report {\it ab initio} electronic structure calculations 
on the different phases of this compound, both within the local spin density approximation (LSDA) and the LSDA+U, in which local electron 
correlation effects are partially accounted for. The results of our calculations clearly show that KCrF$_3$ and LaMnO$_3$ are not only structurally, but also electronically very similar.

\begin{figure}
\centerline{\includegraphics[width=\columnwidth,angle=0]{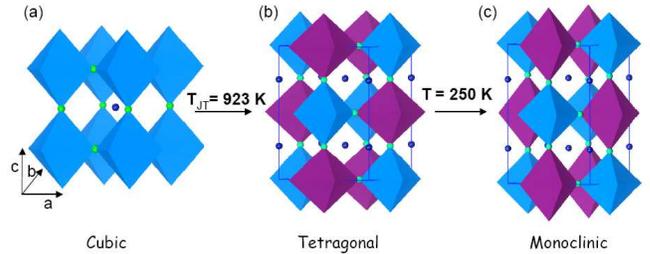}}
\caption{(a) Cubic perovskite crystal structure of KCrF$_3$ at high temperature. (b) Schematic representation of the intermediate temperature tetragonal structure with CrF$_6$ octahedra elongated in an alternate fashion along the (110) (blue) and (1-10) (red) directions. (c) Schematic representation of the low temperature monoclinic structure with CrF$_6$ octahedra elongated along the (1-10) (blue) and (001) (red) directions}
\label{fig0}
\end{figure}

KCrF$_3$ displays three different crystallographic structures, see Fig.~\ref{fig0}. At very high temperatures, the system is a cubic 
perovskite (space group {\it Pm-3m})~\cite{Margadonna2}. Below 973 K, the JT-active high spin Cr$^{2+}$ ion induces a lattice distortion to a
body-centered tetragonal unit cell (space group {\it I4/mcm}), 
isostructural to the Cu$^{2+}$ analogue, KCuF$_3$. In the tetragonal phase, the CrF$_6$ octahedra are 
distorted, leading to short Cr-F bonds along the $c$ axis and alternating long-short Cr-F bonds in the $ab$ plane, indicative of the presence 
of a staggered type of orbital ordering. On cooling below the 250 K, there is a phase transition to a more complicated monoclinic structure with space group {\it I2/m}.

From our density functional calculations we find that the tetragonal phase of KCrF$_3$ is a strongly correlated insulator with a gap of 0.49 eV in LSDA and 2.48 eV in LSDA+U (with $U=6$ eV). For this value of $U$, the calculated relaxed lattice structure is in excellent agreement with the experimental one. In the tetragonal orbitally-ordered phase we find a crystal-field splitting between the Cr $t_{2g}$ and $e_g$ states is 1.0 eV and a total energy gain related to the Jahn-Teller splitting of the $e_g$ states of 0.328 eV per formula unit containing four chromium ions. 
The cooperative Jahn-Teller distortion is accompanied by A-type antiferromagnetic spin ordering in a similar fashion to LaMnO$_3$~\cite{Goodenough63,Solovyev96,Satpathy96,Tyer03}. We find a magnetic moment of $3.85$ {$\mu_B$} per Cr$^{2+}$ ion, in excellent agreement with experiment~\cite{Margadonna3} and  in-plane and out-of-plane superexchange parameters of 2.6 meV and -3.4 meV, respectively. 

The system displays antiferrodistortive ordering of the $3d_{3x^2-r^2}$ and $3d_{3y^2-r^2}$ orbitals in the $ab$ plane --an ordering motif very different from the orbital ordering in KCuF$_3$ that gives rise to a quasi one-dimensional spin chain formation,  and rather resembling the orbital ordering in LaMnO$_3$~\cite{Goodenough63,Solovyev96,Satpathy96,Tyer03,Murakami98,Benedetti01,Liechtenstein95,Caciuffo02}. Along the $c$ axis, the orbital ordering pattern in KCrF$_3$ is rotated by $90$ degrees in consecutive layers. This in contrast to the manganite where the ordering along the $c$ axis is a uniform repetition of the in-plane orbital structure. Another difference with LaMnO$_3$ is that the $e_g$ bandwidth in the chromium compound, as computed within LSDA, is about a factor of two smaller. However, this is compensated in LSDA+U, which shows a bandwidth of the lower Hubbard $e_g$ band of 2.1 eV.

\begin{figure}
\includegraphics[height=\columnwidth,angle=-90]{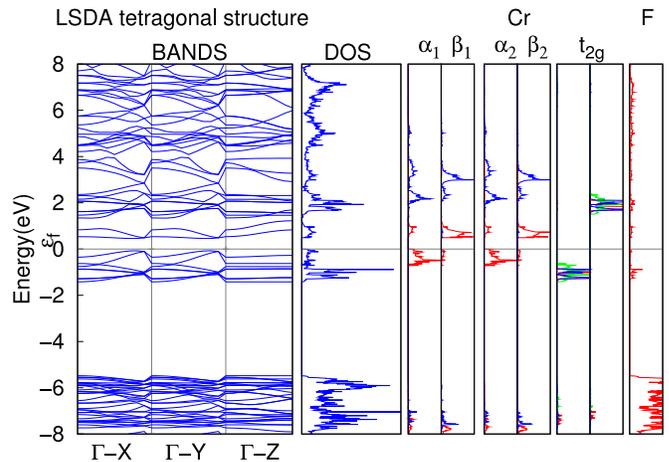}
\caption{Band structure and projected density of states calculated in LSDA for the tetragonal structure of KCrF$_3$. 
The labels ${\alpha}_i$, ${\beta}_i$ are indicating the states ${3x^2-r^2}$, $y^2-z^2$ and ${3y^2-r^2}$, $x^2-z^2$, respectively on neighboring Cr sites $1$ and $2$ in the $ab$ plane.}
\label{ldosLDA}
\label{fig1}
\end{figure}

On cooling below the 250 K, KCrF$_3$ shows a phase transition to a more complicated monoclinic structure (space group {\it I2/m}) with four chromium atoms in the unit cell. 
Our calculations show that in the monoclinic phase an A-type magnetic structure is also realized and that the Jahn-Teller energy is lowered, 
leading to an even stronger orbital ordering. However, the resulting electronic gap and magnetic moment of the compound barely change.

In the following we will present the electronic structure calculations for the three different crystallographic structures. For each one, we considered several possible magnetic ordering structures (ferromagnetic and antiferromagnetic of A, G, and C-type) and analyzed the resulting electronic properties, Jahn-Teller energies and orbital orderings.

\section{Intermediate-Temperature Tetragonal Phase}
The structural changes which occur on lowering the temperature below 973 K through the
cubic-to-tetragonal phase transition can be described in terms of two components: 
a uniform $Q_3$-type tetragonal distortion, which shortens one lattice constant (along the $c$ direction) and lengthens the other two (along 
the $a$ and $b$ directions) and a $Q_2$-type staggered distortion, which introduces alternating Cr-F bond lengths in the $ab$ plane with two distinct 
Cr-F bonds of $2.294/1.986$ \AA. This is a textbook example of  a cooperative Jahn-Teller distortion. The lattice parameters of the resulting 
body-centered tetragonal unit cell at room temperature are $a$= 6.05230 \AA  \ and $c$= 8.02198 \AA~\cite{Margadonna}. 

All our self-consistent calculations are done within LSDA and LSDA+U using the Vienna Ab-initio Simulation 
Package (VASP)\cite{VASP1}. Total energies for the tetragonal structure were calculated with a kinetic cutoff energy of $500$ eV and 
the tetrahedron with Blochl correction, using $105$ irreducible k-points.

\subsection{LSDA Electronic Structure of Tetragonal KCrF$_3$}

We start our study of the tetragonal structure of KCrF$_3$ at the LSDA level and then proceed to also include local correlations within LSDA+U.  
We find that the A-type antiferromagnetic spin ordered structure is the ground state. In Fig. \ref{fig1}, the band structure and the (projected) 
density of states are shown. The system is insulating with an energy gap of $0.49$ eV, which is induced by the Jahn-Teller splitting of 
the $e_g$ states.  In accordance with Hund's rule, the Cr$^{2+}$ ions are in a high spin ${{t_{2g}}^3}{e_g}^{1}$ state, giving rise to a magnetic 
moment of $3.59$ $\mu_B$ per Cr.

\begin{figure}
\includegraphics[height=.6\columnwidth,angle=-90]{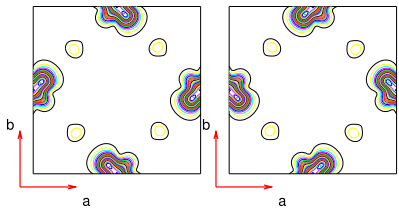}
\caption{Contour plot of charge density corresponding to the occupied $e_g$ bands within the LSDA for the tetragonal structure of KCrF$_3$. The orbital ordering pattern is clearly seen along the bonds connecting the $Cr$ ions, with $3d_{3x^2-r^2}$ and $3d_{3y^2-r^2}$ alternating in the $ab$ plane. The left and right panel are two cuts on consecutive planes along the $c$ direction.
}
\label{fig2}
\end{figure}

The Fermi level lies just above the bands with $t_{2g}$ and $e_g$ character, in agreement with the high spin state of the Cr ions. The exchange splitting 
is about $2.6$ eV, which moves the minority-spin bands far above the Fermi level. The  $t_{2g}$-$e_g$ crystal field splitting, ${\Delta}_{CF}$ is 
about $1.0$ eV.  The occupied Cr bands show little dispersion along the $\Gamma-Z$ direction and are therefore of quasi-two-dimensional character, 
which is due to the specific ordering of the $e_g$ orbitals that maximizes hybridization in-plane and minimizes the out-of-plane dispersion. The 
character of occupied $e_g$ bands is mixed between the two types of $e_g$ states but is mainly coming from ${3x^2-r^2},{3y^2-r^2}$ orbitals on 
neighboring Cr ions. This becomes immediately clear from the contour plot of charge density corresponding to the $e_g$ bands below the Fermi level, 
shown in Fig.~\ref{fig2}. The orbital ordering is found to be staggered along the $c$ direction. The unoccupied orbitals have $3d_{y^2-z^2}$ and 
$3d_{x^2-z^2}$ character, respectively. The  $t_{2g}$ projected density of states resolved for orbital character shows that the $xy$ states have a 
different distribution in energy from the two-fold degenerate orbitals of $yz,zx$ character, in agreement with crystal field symmetry expectations.

The band width of the $t_{2g}$ bands is $0.65$ eV, while the Jahn-Teller split ${e_g}{^1}$ bands just below and above the Fermi level each have a width of about 0.65 eV, almost a factor of two smaller than the value of 1.2 eV in LaMnO$_3$~\cite{Satpathy96}. However, the inclusion of local correlations within LSDA+U changes this bandwidth significantly.

\subsection{LSDA+U Electronic Structure of Tetragonal KCrF$_3$}

\begin{figure}
\includegraphics[height=.5\columnwidth,angle=-90]{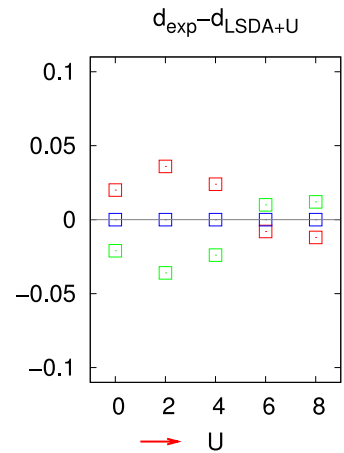}
\caption{Deviation of relaxed Cr-F distances $d1$, $d2$ and $d3$ from their experimental as a function of the Coulomb interaction $U$.}
\label{Fig:fig2_bis}
\end{figure}

It is well known that the incorporation of local Coulomb interactions is essential to understand the physical properties of transition metal 
compounds~\cite{Imada98}. In LSDA+U the electron-electron interaction is dealt with on a meanfield level and we repeated 
the LSDA calculations above within this scheme. 

We performed calculations for a series of values of the on-site Coulomb parameter $U$, namely $U=2.0, 4.0, 6.0, 8.0$ eV, adopting a value for Hund's exchange of $J_H=0.88$ eV. In practice the exact definition of $U$ in a solid is not trivial. The value that is found for this parameter depends on for instance the precise choice of the orbitals that are used in the calculation~\cite{Cococcioni05,Pickett98,Schluter89}. In order to determine its value, we performed a structural optimization as a function of $U$ and subsequently stay with the value for $U$ for which we find an equilibrium structure that matches the experimental one. This scheme to extract the Coulomb parameter is viable because the on-site Hubbard $U$ determines for a large part the orbital polarization of the $e_g$ states, which in turn causes the structural Jahn-Teller lattice distortion~\cite{Liechtenstein95}. 

Hund's exchange parameter $J_H$, in contrast, represents a local multipole and is only very weakly screened in the solid and therefore close to its bare atomic value. 
For it we used the value for a high spin $d^4$ configuration determined by constrained density functional calculations~\cite{Satpathy96}. At any rate small changes of $J_H$ will not affect the results of LSDA+U significantly, as $U$ is the dominating parameter. 

\begin{table}
\centering\begin{tabular}{|c|c|c|c|c|c|c|}
\hline\hline
$\phantom{bl}U \phantom{bl}$ & $\phantom{bl} J_H \phantom{bl}$ & F-type & C-type & G-type & $\phantom{bl}$ $\Delta$ $\phantom{bl}$ & $\phantom{bl}$ $\mu$ $\phantom{bl}$ \\ 
\hline
\hline
0.0 & 0.00 & 0.0103 & 0.4449 & 0.4791 & 0.485  & 3.58 \\
2.0 & 0.88 & 0.0343 & 0.2690 & 0.2476 & 0.810  & 3.63 \\
4.0 & 0.88 & 0.0289 & 0.1303 & 0.1051 & 1.735  & 3.72 \\
6.0 & 0.88 & 0.0206 & 0.0747 & 0.0553 & 2.479  & 3.85 \\
8.0 & 0.88 & 0.0141 & 0.0474 & 0.0339 & 3.332  & 3.98 \\
\hline\hline
\end{tabular}
\caption{Energy difference (eV) between the magnetic ground state (A-type) and other magnetic orderings, band gap $\Delta$ (eV) and 
magnetic moment ($\mu_b$) of the Cr-ions in KCrF$_3$.}
\label{diffenergy}
\end{table}

To determine $U$ we optimized the three inequivalent Cr-F distances in the tetragonal unit cell, $d_1$, $d_2$ and $d_3$, while fixing the lattice parameters $a$, $b$ and $c$ by minimizing the total energy. The results are shown in Fig.~\ref{Fig:fig2_bis}. For $U=6.0$ eV we find that the computed structure is very close to the one obtained experimentally, motivating us to adopt this value as the most reliable one. The obtained $U$ value agrees well with that for Mn$^{4+}$ in LaMnO$_3$ (8.0 eV)~\cite{Satpathy96} calculated within a constrained LDA+U scheme. In KCuF$_3$ a value of 7.5 eV is similarly found for Cu$^{2+}$~\cite{Liechtenstein95}. As one expects that  the ionic core potential of Cr$^{2+}$ causes the $d$ electrons to be less localized with respect to both examples above, it is reasonable to find the smaller value of $U=6.0$ eV  for KCrF$_3$.

\begin{figure}
\includegraphics[height=\columnwidth,angle=-90]{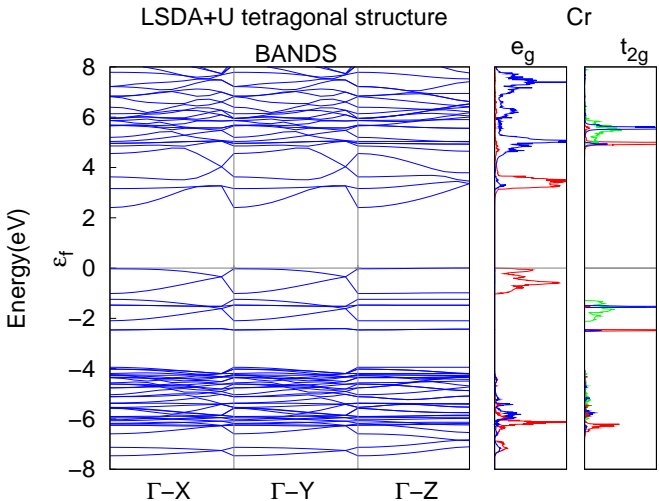}
\caption{Band structure calculated in LSDA+U with $U=6.0$ and $J=0.88$ eV for the tetragonal structure of KCrF$_3$ and (projected) densities of states.}
\label{fig3}
\end{figure}

Below 46 K antiferromagnetic spin ordering is observed~\cite{Margadonna}. To study the magnetic exchange couplings between the Cr ions we calculated the 
total energy of various magnetic structures. The different magnetic structures we considered are: A-type (the spins are parallel in the $ab$ planes 
and the spins are antiparallel along the $c$-axis), F-type (all spins parallel), C-type (each spin is antiparallel to all others in the $ab$ plane 
but parallel along the $c$-axis), and G-type (every spin is antiparallel to all its neighbours).

From the computations we find that the ground state is A-type spin ordered for all values of $U$. The difference in energies between the various magnetically ordered structures for LSDA and LSDA+U is reported in Table~\ref{diffenergy}. From these total energies it is possible to calculate the exchange constants: we find $J_1=2.6$ meV, while $J_2=-3.4$ meV for LSDA+U ($U=6.0$ and $J_H=0.88$). These quantities can be compared with the exchange constants of LaMnO$_3$ \cite{Solovyev96}, where $J_1=9.1$ meV and $J_2=-3.1$ meV.

In Fig. \ref{fig3} the resulting band structure and density of states for the tetragonal structure of KCrF$_3$ within LSDA+U 
($U=6.0$ and $J=0.88$ eV) are shown. The LSDA band gap of  $0.485$ eV increases to a value of 2.479 eV ($U$=6.0 eV), see Table~\ref{diffenergy}.  

From the projected density of states we see that within LSDA+U around the Fermi level there is a clean distribution of the $e_g$ states ${3x^2-r^2}$/${3y^2-r^2}$, depending on the Cr-site. A concomitant enhancement of the orbital polarization is visible in the contour plot of the 
charge density of the occupied $e_g$ bands, see Fig.~\ref{fig4}. This plot also shows that there is an increase in hybrization between the Cr $e_g$ 
states and fluoride $p$ states which enhances the total band width of the occupied Cr $3d$ $e_g$ bands to about $2.1$ eV. The two-dimensional character 
of the occupied $e_g$ bands does not change in the LSDA+U treatment, but the dispersion of the empty $e_g$ states comes to the fore more clearly. 
From the computations on the cubic phase in the next section it will be particularly clear that the two-dimensional character of the occupied $e_g$ 
bands that is caused by the orbital ordering is also the driving force behind the A-type magnetic ordering, as can be expected on the basis of the 
Goodenough-Kanamori~\cite{Goodenough63} rules for superexchange.

\begin{figure}
\includegraphics[height=.9\columnwidth,angle=-90]{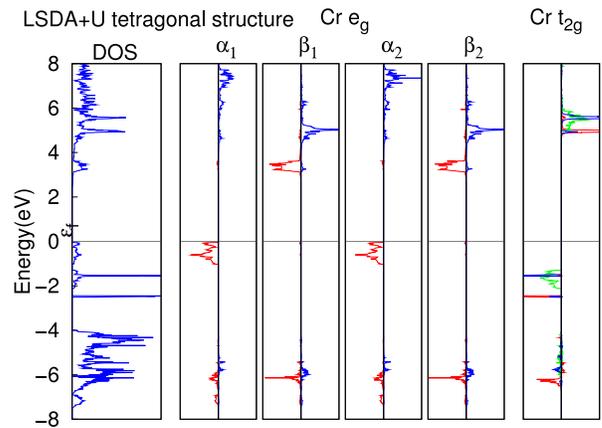}
\caption{DOS projected on different Cr $d$-orbitals, calculated in LSDA+U with $U=6.0$ and $J=0.88$ eV for the tetragonal structure of KCrF$_3$. 
Majority/minority contributions to the DOS are plotted towards the left/right. The labels ${\alpha}_i$, ${\beta}_i$ label the states ${3x^2-r^2}$, $y^2-z^2$ and ${3y^2-r^2}$, $x^2-z^2$ for the different Cr 1 and 2 sites in the $ab$ plane.}
\label{fig3_bis}
\end{figure}

\begin{figure}
\includegraphics[height=.6\columnwidth,angle=-90]{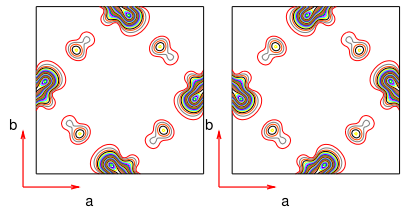}
\caption{Contour plot of charge density corresponding to the occupied $e_g$ bands within LSDA+U for $U=6.0$ and $J=0.88$ eV for the tetragonal structure
of KCrF$_3$.}
\label{fig4}
\end{figure}
 
\section{High-Temperature Cubic Phase}
In the cubic $Pm-3m$ structure ($a= 4.231783$ \AA)~\cite{Margadonna2}, the distances between all Cr and neighboring F-ions are equal to 2.116 \AA \ and the $e_g$ states are locally degenerate.  We find that at the LSDA level, cubic KCrF$_3$ is metallic, for the groundstate A-type magnetic structure. Such is expected as in the absence of a Jahn-Teller distortion the $e_g$ band is half filled, even though it is fully spin polarized. 

\begin{figure}
\includegraphics[height=\columnwidth,angle=-90]{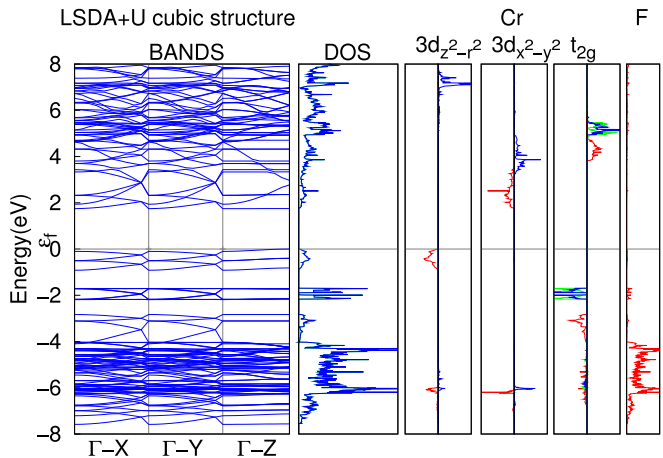}
\caption{LSDA+U ($U$=6.0 and $J_H$=0.88 eV) band structure and projected DOS for cubic KCrF$_3$ with $Pm-3m$ symmetry and A-type magnetic ordering.}
\label{fig7}
\end{figure}

In LSDA+U a bandgap of 1.72 eV opens up --the band structures calculated by LSDA+U ($U=6.0$ and $J=0.88$ eV) with the cubic unit cell with A-type magnetic ordering is shown in~\ref{fig7}. The correlation-induced Mott gap is smaller than the charge gap in the tetragonal structure because of the absence of the Jahn-Teller distortion.

\begin{figure}
\includegraphics[height=.6\columnwidth,angle=-90]{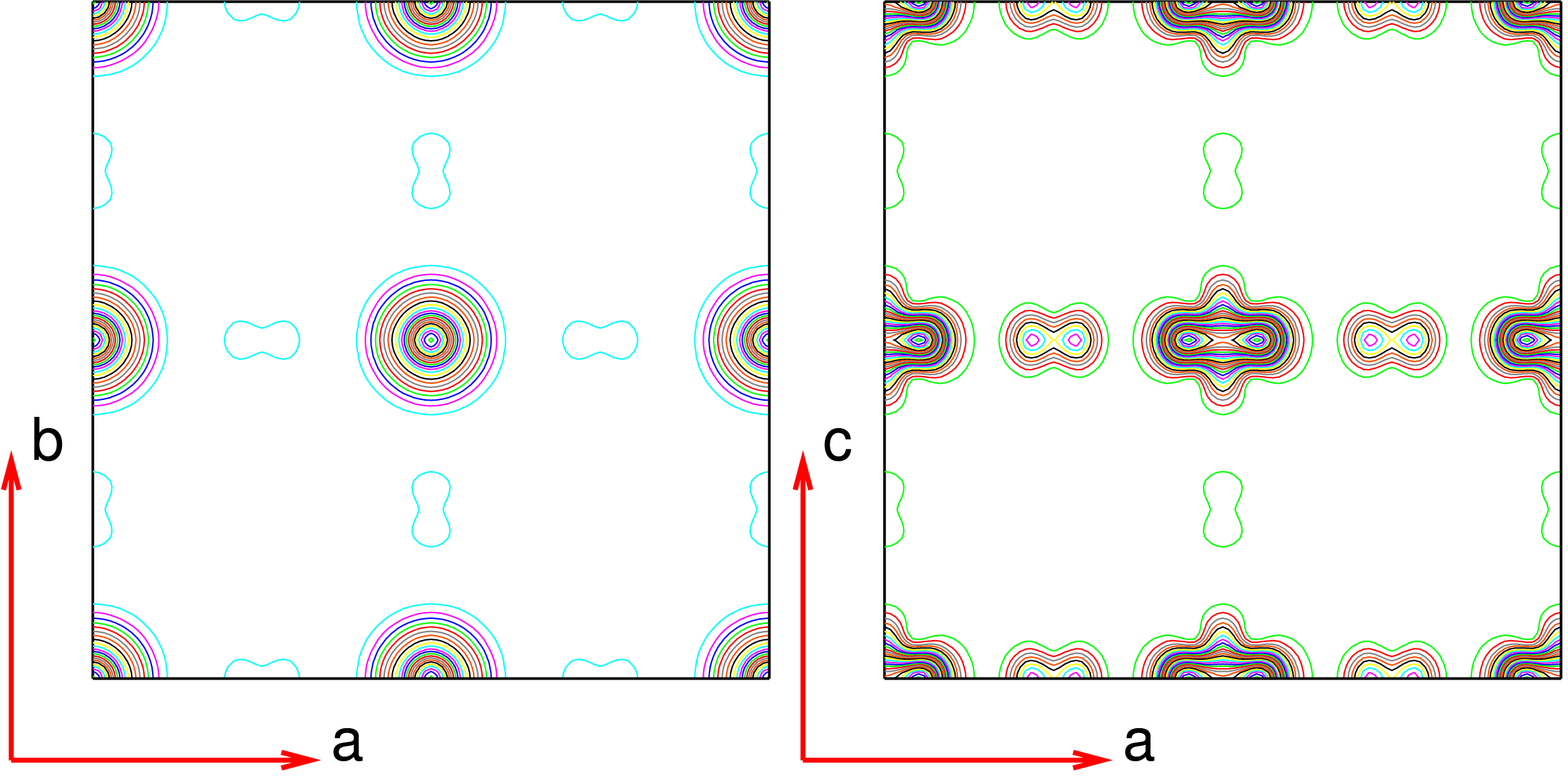}
\hspace{1cm}
\includegraphics[height=.6\columnwidth,angle=-90]{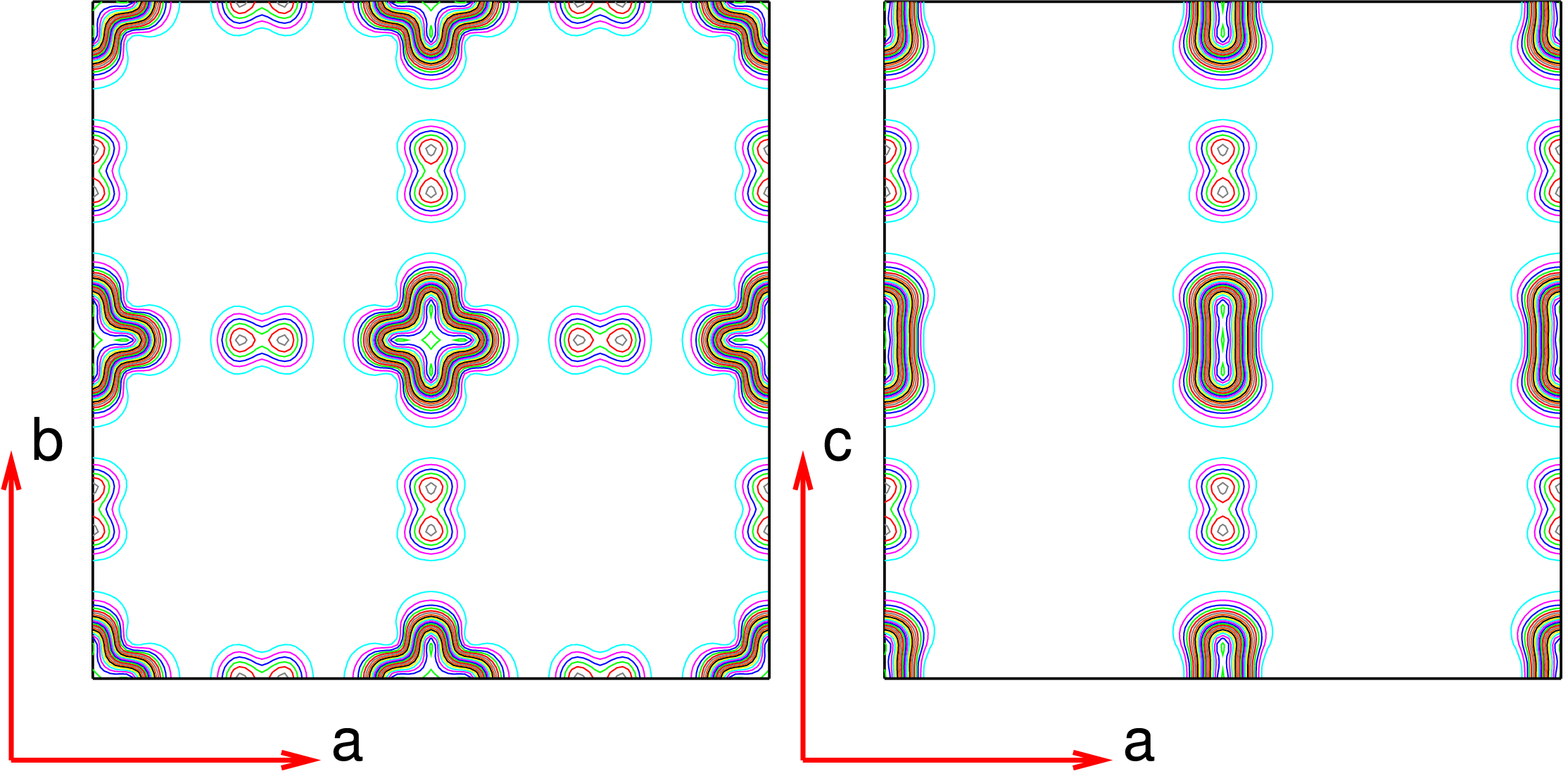}
\caption{Orbitals in the cubic phase of KCrF$_3$, obtained with LSDA+U. Left: orbital ordering of $x^2-y^2$ orbitals for C-type spin ordering. Right: orbital ordering of $3z^2-r^2$ orbitals for A-type spin ordering. Top panels: side view. Bottom panels: top view}
\label{fig8}
\end{figure}

Despite the fact that the Jahn-Teller distortions are absent in this structure, there is still an orbital ordering which is due to the magnetic exchange, see Fig.~\ref{fig8}. 
This exchange-driven orbital ordering can be understood in terms of the orbital dependence of the superexchange energy between neighboring Cr-sites. Such a 
situation is described in terms of a Kugel-Khomskii model~\cite{Kugel82}. For the A-type magnetic ordering we obtain a homogeneous orbital occupation of $3z^2-r^2$ states, oriented perpendicular to the ferromagnetic plane. This is in accordance with the Goodenough-Kanamori~\cite{Goodenough63} rules for superexchange: bonds of occupied $3z^2-r^2$ orbitals on top of each other have a large overlap and therefore result in antiferromagnetic spin ordering. Within the plane the overlap is mainly between occupied $3z^2-r^2$ and empty $x^2-y^2$ orbitals, causing a ferromagnetic orientation of the spins.

When we consider C-type magnetic ordering the resulting orbital ordering (see Fig.~\ref{fig8}) is of the homogeneous $x^2-y^2$ type --again with 
antiferromagnetic spin orientation for orbitals with lobes pointing towards each other ($x^2-y^2$ orbitals in the plane) and ferromagnetic 
orientation between orbitals with small overlap. 

The A- and C-type order that we considered are just two of many possible magnetic orderings with 
concomitant orbital orderings: other configurations can appear when the unit cell is doubled or quadrupled in accordance with model calculations on 
the Kugel-Khomskii Hamiltonian~\cite{Kugel82,Khaliullin97,Brink99b,Brink04b}. These observations of magnetically driven orbital ordering 
(or orbitally driven magnetic ordering, depending on one's point of view), although interesting from a theoretical perspective, are not directly relevant to the experimental situation as at the high temperatures where the cubic phase is stable no long range magnetic ordering is expected.

By comparing the total energies of the cubic and tetragonal phases in the paramagnetic state we can directly compute the energy gain in the tetragonal 
phase that is due to the Jahn-Teller distortion alone.  
We obtain a value of ${{\Delta}E}_{JT}=0.328$ eV per unit formula, which is comparable to that found in LaMnO$_3$ (0.504 eV)~\cite{Tyer03}.

\section{Low-Temperature Monoclinic Phase}
Below 250 K, KCrF$_3$ shows a phase transition to a monoclinic structure, characterized by a pronounced tilting of the CrF$_6$ octahedra. 
The lattice parameters at 150 K are $a= 5.82642$ \AA, $b= 5.83517$ \AA, $c= 8.57547$ \AA \ and $\gamma=93.686^o$~\cite{Margadonna}. We performed the total energies calculations with a kinetic cutoff energy of $500$ eV and used the tetrahedron method with Blochl correction, using $90$ irreducible k-points. 

The total energy that we compute in the monoclinic phase reveals that the Jahn-Teller distortion is further stabilized with an energy gain of $22$ meV per Cr with respect to the tetragonal phase. Again in the LSDA+U calculations the magnetic groundstate is found to be A-type, the Cr moment is $3.85 {\mu_b}$, the band gap $2.49$ eV and the orbital ordering is essentially the same as in the tetragonal phase (see Fig.~\ref{fig10}). Per unit formula the ferromagnetic configuration is higher in energy by
$0.0168$ eV, the C-type configuration by $0.0438$ eV and the G-type state by $0.0275$ eV.

From this we find slightly larger in-plane and inter-plane magnetic couplings of $J_1=2.1$ meV and $J_2=-1.7$ meV, respectively. Similar to LaMnO$_3$\cite{Skmryev99}, experimentally a small spin canting is observed, giving rise to weak ferromagnetism in the monoclinic phase~\cite{Margadonna}. When the CuF$_6$ octahedra are tilted the weak local anisotropy and non-local Dzyaloshinskii-Moriya interaction lead to spin canting. 
In a first principles bandstructure calculation, this can be taken into account when on top of the present computation scheme the relativistic spin-orbit interaction is included~\cite{Solovyev96}.

\begin{figure}
\includegraphics[height=\columnwidth,angle=270]{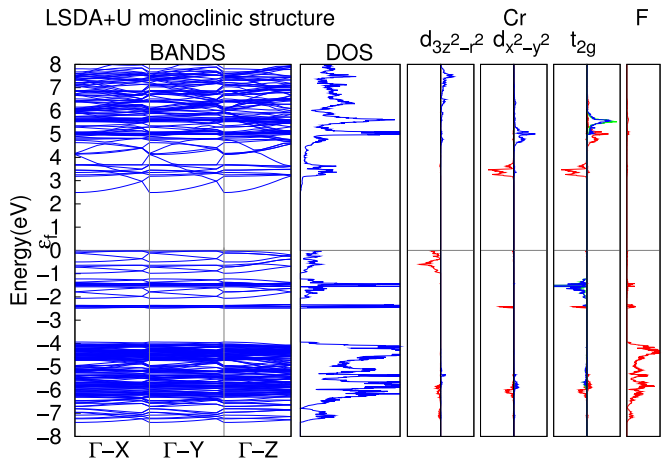}
\caption{Monoclinic phase of KCrF$_3$: band structure and DOS calculated in LSDA+U with $U=6.0$ and $J=0.88$ eV. }
\label{fig9}
\end{figure}

\section{Conclusions}
With a set of density functional calculations we determine the electronic, magnetic and orbital properties of KCrF$_3$. Our ternary chromium fluoride shows 
many similarities with LaMnO$_3$. From the electronic point of view, the bandgap and conduction bandwidth are comparable, although somewhat smaller 
for the more ionic KCrF$_3$ in the orbitally ordered phase. The magnetic structure is of the same A-type and the exchange constants are of the same order of magnitude. The orbital ordering in the ferromagnetic planes is identical in the two compounds, although the stacking of the ordering along the $c$-axis is different. These properties of KCrF$_3$ make it a material that is comparable to LaMnO$_3$ and therefore attractive to investigate for instance orbital excitations~\cite{Saitoh01,Brink99b,Perebeinos00,Brink01} and orbital scattering in photoemission~\cite{Brink00}. Doping the strongly correlated Mott insulator KCrF$_3$ with electrons or holes may lead to very interesting prospects, as the equivalent manganites show an overwhelming wealth in physical properties. If the concentration and kinetic energy of the doped carriers suffices a melting of orbital ordering is anticipated, establishing an orbital liquid phase, changing the electronic dimensionality from effectively two to three~\cite{Brink04a,Brink06}. In the manganites a colossal magneto-resistance is observed in the vicinity of such a phase transition.

\acknowledgments

This work was financially supported by ``NanoNed'', a nanotechnology
programme of the Dutch Ministry of Economic Affairs and by the
``Nederlandse Organisatie voor Wetenschappelijk Onderzoek (NWO)''
and the ``Stichting voor Fundamenteel Onderzoek der Materie (FOM)''.
Part of the calculations were performed with a grant of computer
time from the ``Stichting Nationale Computerfaciliteiten (NCF)''.

\begin{figure}
\includegraphics[height=0.4\columnwidth,angle=-90]{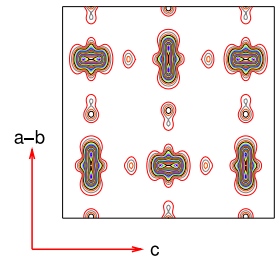}
\caption{Contour plot of charge density corresponding to the occupied $e_g$ bands below the Fermi level for the monoclinic structure of KCrF$_3$ in LSDA+U.}
\label{fig10}
\end{figure}

\end{document}